\RequirePackage{ifpdf}
\ifpdf 
\documentclass[pdftex]{sigma}
\else
\documentclass{sigma}
\fi

\newcommand{\bbc}{{\mathbb C}}

\newcommand{\bbi}{{\mathbb I}}
\newcommand{\bbj}{{\mathbb J}}

\newcommand{\bbp}{{\mathbb P}}

\newcommand{\bbs}{{\mathbb S}}

\newcommand{\bbz}{{\mathbb Z}}

\newcommand{\Got}[1]{{\mathfrak #1}}

\newcommand{\gh}{{\Got h}}

\begin{document}

\allowdisplaybreaks

\renewcommand{\PaperNumber}{017}

\renewcommand{\thefootnote}{$\star$}

\FirstPageHeading

\ShortArticleName{Separation of Variables and the Geometry of
Jacobians}

\ArticleName{Separation of Variables\\ and the Geometry of
Jacobians\footnote{This paper is a contribution to the Vadim
Kuznetsov Memorial Issue ``Integrable Systems and Related
Topics''. The full collection is available at
\href{http://www.emis.de/journals/SIGMA/kuznetsov.html}{http://www.emis.de/journals/SIGMA/kuznetsov.html}}}

\Author{Jacques HURTUBISE}

\AuthorNameForHeading{J. Hurtubise}

\Address{Department of Mathematics and Statistics, McGill University,\\
 805 Sherbrooke St. W. Montreal H3A 2K6, Canada}
 \Email{\href{mailto:jacques.hurtubise@mcgill.ca}{jacques.hurtubise@mcgill.ca}}
\URLaddress{\url{http://www.math.mcgill.ca/hurtubise/}}

\ArticleDates{Received November 17, 2006, in f\/inal form January
08, 2007; Published online February 05, 2007}

\Abstract{This survey examines  separation of variables for
algebraically integrable Hamiltonian systems whose tori are
Jacobians of Riemann surfaces. For these cases there is a~natural
class of systems which admit separations in a nice geometric
sense. This class includes many of the well-known cases.}

\Keywords{separation of variables; integrable Hamiltonian systems;
geometry of Jacobians}

\Classification{37J35; 70H06} 

\section{Separation of variables and complex integrable systems}

The general subject of separation of variables is one with a long
history and a wide variety of techniques, but, to a geometer at
least, it is  somewhat puzzling, if one is trying to f\/ind some
geometric context which explains it. It does turn out, however,
that from the viewpoint of complex geometry, there is a quite
specif\/ic context in which there is some   beautiful geometric
content, and this is the subject of this survey. Over the years, I
had the pleasure of discussing these questions several times with
Vadim Kuznetsov, and always prof\/ited from his insights. It is an
honour to be able to dedicate this article to his memory.

Separation of variables for real Hamiltonian integrable systems
is, in some sense, automatic, at least if one is  dealing with
systems with compact level sets for the Hamiltonians and if one
can reparametrize the ring of Hamiltonians: the existence of
action-angle variables ensures that the Hamiltonian f\/lows can be
decoupled into one-dimensional problems. The level sets of the
Hamiltonians are tori, and the angle coordinates express these as
products of circles. The problem, of course, is not one of knowing
that it can be done, but rather of doing it, in some fairly
explicit way, on some explicit family of examples. Over the
complex numbers, however, we shall see that the question of
whether one can separate variables in some reasonable sense
becomes tied in with the actual geometry of the level sets; the
abstract existence of a separation becomes linked to its
ef\/fective computation.

One can imagine starting with some real analytic integrable
system, and complexifying it. A~f\/irst question is whether the
level sets of the real system, which we will assume to be compact
and so, generically, tori, remain so once complexif\/ied,
compactif\/ied, and eventually desingularised. This is far from
being automatic; the passage to the complex domain reveals more
elaborate geometry: one need only think of algebraic curves in the
plane, which are unions of ovals; their complexif\/ications reveal
much more intricate structure. Nevertheless, there are real
integrable systems which complexify well, in the sense that the
level sets of the Hamiltonians are complex tori, that is, real
tori topologically, but with a complex structure; we refer the
reader to the paper \cite{AvM2} and the references therein for a
discussion of this question of whether one still has tori over the
complex domain, which is in fact linked, in the case when the tori
are in addition algebraic (Abelian varieties),  to the existence
of large families of meromorphic local solutions, i.e., to some
sort of Painlev\'e test.

We will assume that our integrable system, now complex of
dimension $2r$, has as Lagrangian leaves (level sets of the
Hamiltonians) a family of open dense subsets of complex tori of
complex dimension $r$, at least over some dense open set: the
leaves compactify  to give complex tori. In the examples, our tori
are algebraic, and the systems are what is known as algebraically
integrable.

Over the complex numbers, the tori are far from having a constant
structure. One can, as in the real case, obtain action-angle
variables $I_{j}, \theta_{j}$, so that the symplectic form is
given by $\sum_{j} dI_{j} \wedge d\theta_{j}$ and the Lagrangian
leaves ${\mathcal L}_{c}$, $c=(c_{1},\dots,c_{r})\in \bbc^r$ are
cut out by $I_{j}= c_{j}$. The tori are of the form
\begin{gather*}
{\mathcal L}_{c}= \bbc^{r}/(\bbz^{r} + \Lambda(c)\cdot\bbz^{r}),
\end{gather*}
where $\Lambda(I) $ is some matrix-valued function of the $I_{j}$
with non-degenerate imaginary part. One can show that for there to
be a symplectic form on the f\/ibration of tori such that the
individual tori are Lagrangian, one needs
\begin{gather*}
\frac{\partial\Lambda_{ij}} {\partial I_{k}}=
\frac{\partial\Lambda_{ik}} {\partial I_{j}}
\end{gather*}
so that $\Lambda_{ij} = {\partial G_{i}}/{\partial I_{j}}$ for
some $G_{i}(I)$.  This is the explicit version of the `cubic
condition' of Donagi--Markman \cite
{donagi-markman-cy-threefolds}. It is simply a consequence of the
need for the form $\sum_{j} dI_{j} \wedge d\theta_{j}$ to be
invariant under translation by the lattice
$\Lambda(I)\cdot\bbz^{r}$.

Thus the functions $G_{i}(I)$ distinguish algebraically integrable
systems, essentially by their derivatives being parameters for
moduli of the tori in the f\/ibers; one can derive invariants from
them. We are interested in cases for which the tori are in some
reasonable sense symmetric products of one dimensional compact
manifolds, as this will be necessary for separation of variables
in the integrable system.

 {\samepage
 A natural class for which this occurs are the Jacobians $J(S)$ of
compact Riemann surfaces~$S$. Indeed, if $S$ is such a surface, of
genus $g$, one has a $g$-dimensional space of holomorphic
one-forms on $S$; let $\omega_{i}$, $i = 1,\ldots,g$ be a basis
for this space. The integral of any one-form along~$S$ from one
point to another is only def\/ined modulo periods. The f\/irst
homology of $S$ is $\bbz^{2g}$; let $c_{j}$ be a~(standard,
symplectic) basis for this group. One can normalise so that the
periods $\oint_{c_{j}} \omega_{i}= \delta_{ij}$, $i, j \in
\{1,\ldots,g\}$; let $\Lambda_{ij}= \oint_{c_{g+j}}\omega_{i}$
generate the remaining periods. We set
\begin{gather*}
J(S) = \bbc^{g}/(\bbz^{g}+\Lambda\cdot\bbz^{g}).
\end{gather*}
Choosing a base point $p_{0}$, one has the Abel--Jacobi  map
\begin{gather*}
A\colon S \rightarrow  J(S),\qquad p \mapsto \int_{p_{0}}^{p}
\omega_{1},\ldots,\omega_{g}. \nonumber
\end{gather*}
One can extend this map to the symmetric products of $S$:
\begin{gather*}
A^{(r)}\colon S^{r}  \rightarrow J(S),\qquad (p_{1},\ldots,p_{r})
\mapsto  \sum_{i=1}^{r}\int_{p_{0}}^{p_{i}}
\omega_{1},\ldots,\omega_{g}.
\end{gather*}
The basic fact that drives a lot of what follows is the fact, due
to Jacobi, that the map
\begin{gather*}
A^{(g)}\colon S^{g} \rightarrow J(S)
\end{gather*}
is onto and  generically bijective.

}

In the case when the Lagrangian tori of the integrable systems are
Jacobians, then, we are half way there in terms of having a
separation of variables, in that we are decomposing the Jacobian
as a symmetric product of the curve. The purpose of this note is
to highlight some results concerning algebraically integrable
systems for which a separation of variables of the whole phase
space in terms of symmetric products is natural, extending the
geometry of the Jacobians; as we are dealing with phase spaces,
the symmetric product is of a surface, which should contain the
curves in our family. Two sets of $g$ points will belong to the
same Lagrangian leaf if they belong to the same curve in the
family.  The separation in these coordinates follows easily. The
cases covered include many of the frequently studied examples. In
the sections which follow, we will give a survey, f\/irst of
results concerning the local geometry of integrable systems of
Jacobians, showing that if suitable conditions are met they can be
related to symmetric products of surfaces, or more accurately to
Hilbert schemes of points. After that we discuss some important
cases, covering the various Gaudin models, the Sklyanin brackets
and their generalisations, such as generalised Hitchin systems. We
will also discuss relations to multi-Hamiltonian structures, and
close with a discussion of what happens when one replaces
Jacobians by Prym varieties.

What follows is a survey of results proven elsewhere, in
collaboration with Malcolm Adams, John Harnad,  Mounia Kjiri and
Eyal Markman \cite{AHH2, HaH-multihamiltonian, HK-generalised,
hurtubise-surfaces, HM-sklyanin, HM-pryms}. In relation to the
Gaudin model, a simple version also appears in
\cite{Novikov-Veselov}. Another place in which similar
computations can be found is \cite{GNR}. In computational form,
some cases of the results presented here have a long history: one
already has the result for the rational Gaudin model  in
\cite{Ga}, which appeared in~1919!

\section{The local geometry of systems of Jacobians}

Let $U$ be some $g$-dimensional variety; as we are talking about
local geometry, the typical $U$ would be some small open ball in
$\bbc^{g}$. Let
\begin{gather*}
\bbj\rightarrow U
\end{gather*}
be an integrable system of Jacobians, that is a symplectic $2g$
dimensional variety f\/ibering over $U$, such that the f\/ibers
$J(S_h)$, $h\in U$ are Lagrangian and are  Jacobians $J(S_{h})$ of
genus $g$ Riemann surfaces $S_{h}$. There is an associated family
of Riemann surfaces
\begin{gather*}
\bbs\rightarrow U,
\end{gather*}
with f\/ibers $S_{h}$ over $h\in U$. If we choose base sections,
we have a f\/iberwise Abel map
\begin{gather}
A\colon \bbs\rightarrow \bbj \label{Abel-on-fibration}
\end{gather}
which commutes with the projections to $U$.
 We note that the f\/iberwise Abel map (\ref{Abel-on-fibration}) extends
 to the f\/iberwise symmetric product
 \begin{gather*}
 A^{(g)}\colon \bbs_{U}^{(g)}\rightarrow \bbj.
 \end{gather*}
 Extending the corresponding results for individual Jacobians,
 $A^{(g)}$ is
 onto and generically bijective.

 We can pull back the symplectic form $\Omega$ from $\bbj$ to $\bbs$.
 One has that $A^*\Omega$ never vanishes, essentially because the Abel
 map is an immersion. Its square, however, can, and we say:

 \begin{definition} The system has \emph{rank two} if for a suitable choice of base sections
 the pull-back of the symplectic form satisf\/ies
 \begin{gather*}
A^*\Omega \wedge A^*\Omega =  0.
\end{gather*}
  \end{definition}
  A classical theorem of Darboux tells us that there are then
  coordinates $x, y$ such that \mbox{$A^*\Omega = dx\wedge dy$}; locally the
  form is pulled back from a surface.

 We digress for a moment to discuss Hilbert schemes. In general, these
 parame\-trise subvarieties of an algebraic variety, or more properly
 subschemes (which allows for degeneracy, multiplicity and so on).
 The case we need is associated to a smooth surface $Q$. We denote by
${\rm Hilb}^g(Q )$ the Hilbert scheme which parametrises
0-dimensional
 schemes~$D$ of $Q$ of length $g$. These  schemes~$D$ are generically unions of $g$
 unordered points of $Q$, with more complicated structures when
 points coalesce. The space ${\rm Hilb}^g(Q )$ has a support map ${\rm Hilb}^g(Q )\rightarrow Q^{(g)}$
 to the $g$-th symmetric product, which is an isomorphism over the set
 of $g$-tuples of distinct points. ${\rm Hilb}^{g}(Q)$ is smooth, and is a
 desingularisation of $Q^{(g)}$. When $Q$ is symplectic, or Poisson,
 ${\rm Hilb}^{g}(Q)$  is also symplectic, or Poisson, and the structure is a
 desingularisation of the natural one on $Q^{(g)}$.
 We note that more properly we should be talking of Douady spaces,
 as the surfaces we have are not apriori algebraic. Nevertheless,
 the local structure is identical to that of the well studied Hilbert
 schemes of points on a smooth algebraic surface, and we will keep the notation ${\rm Hilb}^g(Q )$.

\begin{theorem}[\cite{hurtubise-surfaces}]\label{theorem1}
\emph{(i)} Let the system be of rank $2$. Under the embedding $A$,
the variety ${\bbs}$ is coisotropic. Quotienting by the null
foliation, one obtains, restricting $U$ if necessary, a
surface~$Q$ and a projection $\pi\colon  \bbs\rightarrow Q$, such
that $A^*\Omega$ projects, defining a symplectic form $\omega$ on
$Q$ with $\pi^{*}(\omega) = A^{*}(\Omega)$. The Riemann surfaces
$S_h$ all embed in $Q$.

\emph{(ii)}  (independence of base section) If $A$, $\tilde{A}$
are two Abel maps with  $ A^*\Omega \wedge A^*\Omega = 0$, $\tilde
{A}^*\Omega\wedge \tilde {A}^*\Omega = 0$, then
$A^*\Omega=\tilde{A}^*\Omega$, when $g \ge 3$, and so $Q$ depends
only on $\bbs$ and not on the particular Abel map chosen. For $g
=2$,
 $ A^*\Omega \wedge A^*\Omega $ is always zero.

\emph{(iii)}  There is a diagram of maps:
\begin{gather*}
\begin{matrix}
    \bbs_{U}^{(g)}&\rightarrow&\bbj\cr\cr
\downarrow&\nearrow \Phi&\cr\cr {\rm Hilb}^g(Q )&&
\end{matrix}
\end{gather*}
which defines the birational symplectic morphism $\Phi$, an
isomorphism over Zariski open sets, between $\bbj$ and   ${\rm
Hilb}^g(Q )$. The image of symmetric products $S_h^{(g)}$ of the
curves is Lagrangian in ${\rm Hilb}^g(Q)$, and the restriction of
$\Phi$ to $S_h^{(g)}$  is the Abel map
\begin{gather*}
S_h^{(g)}\to J(S_{h}).
\end{gather*}
\end{theorem}

What this tells us, roughly,  is that the surface $Q$ and the
$g$-dimensional family of curves $S_{h}$, $h\in U$ embedded in it
encodes the integrable system. The inclusion of the curve $S_{h}$
in $Q$ is for dimensional reasons Lagrangian, and the induced
inclusion of symmetric products
\begin{gather*}
S_h^{(g)}\rightarrow Q^{(g)},
\end{gather*}
which lifts to the Hilbert scheme,
\begin{gather*}
S_h^{(g)}\rightarrow {\rm Hilb}^g(Q),
\end{gather*}
is still Lagrangian, and def\/ines the Lagrangian leaves of the
integrable system. The isomorphism~$\Phi$ of Theorem
\ref{theorem1}  is, in ef\/fect, the change of variables to
separating coordinates.

The key constraint, of course, is the rank 2 condition.  A related
construction, due to Krichever and Phong \cite{KP} uses the
existence of a meromorphic one-form, whose exterior derivative is
the symplectic form.

\section[$r$-matrix systems and their generalisations]{$\boldsymbol{r}$-matrix systems and their generalisations}

We now outline how a large family of well-studied integrable
systems satisfy the rank 2 condition and therefore admit algebraic
separation of variables. The list of systems comprises all of
those def\/ined using loop algebra-valued $r$-matrices, for the
loop algebra $gl(r)$:
\begin{itemize}\itemsep=0pt
\item the rational $gl(r)$-Gaudin model and its degenerations
\cite {AHH2,HaH-tops}; \item the rational $Gl(r)$ Sklyanin
brackets (magnetic chain) \cite{scott, sklyanin, sklyanin2}; \item
the elliptic and trigonometric $gl(r)$ Gaudin models \cite
{HK-generalised}; \item the elliptic and trigonometric $Gl(r)$
Sklyanin brackets \cite { HM-sklyanin}; \item the $Gl(r)$-cases of
the systems def\/ined using the `dynamical $r$-matrix' formalism
of Felder, Etingof--Varchenko
\cite{etingof-schiffman,etingof-varchenko,  felder,
HM-sklyanin-general}; \item the $gl(r)$ Hitchin systems, and their
generalisations \cite{Bo1, hitchin-self-duality,
hitchin-integrable-system, hurtubise-surfaces, HK-generalised, M}.

\end {itemize}
These systems contain many, if not most,  of the examples of
integrable systems of Jacobians that have been studied (see, for
example, \cite {AHH, AHP, AvM, AvM1, DM, reiman-semenov,
reiman-semenov2}) . Their generalisations to structure groups
other than $Gl(r)$ are also very important, for example with the
orthogonal groups in the study of tops; the Abelian varieties
which occur naturally are not Jacobians, however, but a more
general class known as Prym varieties. We shall discuss these
brief\/ly later.

We f\/irst outline how these systems are indeed integrable systems
of Jacobians. Their phase spaces are all, (after quotienting out,
in some cases, by groups of global automorphisms) spaces of pairs
$(E,\phi)$, where $E$ is a vector bundle over a Riemann surface
$\Sigma$ and $\phi$ is a meromorphic 1-form-valued section of  the
endomorphism bundle $\mathop{\rm End}(E)$ with poles over a
positive divisor $D$:
\begin{gather*}
\phi\in H^{0}(\Sigma, \mathop{\rm End}(E)\otimes K_{\Sigma}(D)).
\end{gather*}
For example, for the (reduced) rational Gaudin model, the  Riemann
surface is the projective line, the bundle $E$ is trivial, and so
$\phi$ becomes a matrix valued meromorphic function
 with poles at $D$, considered modulo conjugation by a constant matrix.

Now let $X$ be the total space of the line bundle $K(D)$ over
$\Sigma$ of one-forms with poles at $D$. One has a projection
$\pi\colon  X\rightarrow  \Sigma$. To each pair $(E,\phi)$ one can
associate a pair $(S,L)$. Here $S$ is the spectral curve of
$\phi$, embedded in $X$, where it is cut out by the equation
\begin{gather}
\det(\phi-\xi\bbi)=0, \label {spectral curve}
\end{gather}
where $\xi$ is the tautological section of $\pi^*K(D)$ over $X$.
$L$, in turn, is the sheaf def\/ined over $X$ but supported over
$S$ given as the cokernel of $\phi$:
\begin{gather*}
0\rightarrow \pi^{*}E\otimes \pi^*K^{*}(-D) \buildrel
{\phi-\xi\bbi}\over{\longrightarrow}\pi^{*}E \rightarrow
L\rightarrow 0.
\end{gather*}
One can recover the pair $(E,\phi)$ from $(S,L)$ as direct images,
$E$ as the direct image of $L$, and $\phi$ as the direct image of
multiplication by the tautological section:
\begin{gather*}
E =\pi_{*}(L),
\\
\phi =\pi_{*}(\times \xi).
\end{gather*}

 One has that over the generic locus of smooth $S$, the sheaf $L$ is
 a line bundle over $S$, and is represented by a point in the
 Jacobian of $S$.

The nice thing is the map which associates to $(S,L)$ the pair
$(E,\phi)$ is Poisson. Indeed, the surface $X$ has a Poisson
structure(indeed, frequently, several), and it is the discovery of
Mukai, Bottacin and Tyurin \cite{Bo2,Mu,Ty} that this structure
induces a Poisson structure on the variety of pairs $(S, L)$. The
family of curves $S$ is a linear system, and is such that
deformations of the curve which f\/ix the intersection with the
degeneracy divisor $D$ of the Poisson structure are given
inf\/initesimally by sections of the canonical bundle of the
curve; the  line bundles are of f\/ixed degree, and their
deformations are given  by the f\/irst cohomology of the structure
sheaf of the curves. The Poisson structure is then derived from
the Serre pairing of these two deformation spaces. We have, in all
the cases quoted above (precise statements for the individual
cases can be found in \cite{AHH2, HaH-multihamiltonian,
HK-generalised, hurtubise-surfaces, HM-sklyanin, HM-pryms}):
 \begin{proposition}
 For a suitable choice of Poisson structure on $X$, the map
 \begin{gather*}
 (S,L)\mapsto(E,\phi)
 \end{gather*}
 is Poisson.
 \end{proposition}

 Viewed invariantly, the result is not dif\/f\/icult, and just expresses
 an invariance under `push-down'. See \cite {HaH-multihamiltonian}.

 The integrable systems under study all have as Hamiltonians the
 coef\/f\/icients of the spectral curve (\ref {spectral curve}). One then
 shows:

 \begin{proposition}
 The fibration
 \begin{gather*}
 (S,L)\mapsto S
 \end{gather*}
 is Lagrangian, with fiber over $S$ the Jacobian of $S$ when $S$ is smooth.
 \end{proposition}

 Finally, by tensoring $L$ by a f\/ixed line bundle if necessary (which
 is always possible locally), we can
 suppose that the degree of $L$ is the genus $g$ of $S$. Generically,
 then, $L$ has only one section $s_{L}$ up to scale.  Let $D_{L}$ be its
 divisor on $S$. $D_{L} $ has degree $g$, and is, generically, the
 union of $g$ distinct points of $S\subset X$. In any case, the
 divisor $D_{L}$ is cut out on $X$  by the equation of the spectral
 curve, and some additional equations, and so can be thought of as an element of ${\rm Hilb}^{g}(X)$.
 A fairly straightforward
 computation shows that over the set where $L$ has a unique section up to scale:

 \begin{proposition}
 The map
 \begin{gather*}
 (S,L)\mapsto D_{L}
 \end{gather*}
 is Poisson.
 \end{proposition}

 This, in essence, shows us that $X$ should be the variety $Q$ of the
 main theorem of the preceding section. This is not quite the case;
 $X$ is only Poisson, while $Q$ is symplectic. One f\/inds, however,
 that the symplectic leaves of the original space of pairs $(E,\phi)$
 give curves which intersect the degeneracy locus $V$ of the Poisson
 structure on $X$ at f\/ixed points. Blowing up $X$ at these points
 gives our variety $Q$.

 It should be emphasised that this correspondence between $(E,\phi)$
 and a set of points on $X$, while presented here in an abstract way,
 is actually fairly ef\/fective computationally, so that if the
 original phase
 space can be described ef\/fectively, then so can the passage to $g$
 points on $X$, and then the integration of the system. We illustrate
 this  with one example, following from \cite {HM-sklyanin}; the general
 procedure is the same in all cases.

 The example in question is that of  the $Gl(r,\bbc)$ elliptic Sklyanin or
 quadratic $r$-matrix bracket. Let $\Sigma$ be an elliptic curve,
 \begin{gather*}
\Sigma = \bbc/(\omega_1\bbz+\omega_2\bbz),
\end{gather*}
 and let $\rho\colon \bbc\rightarrow \Sigma$ be the natural projection. Let
 $q={\rm exp} (2\pi i/r)$,  and def\/ine two $r\times r$ matrices
\begin{gather*}
I_1 = {\rm diag} (1,q, q^2,\dots,q^{r-1}), \qquad I_2 =
\begin{pmatrix} 0&1&0&\cdots&0\cr 0&0&1&\cdots&0\cr \cdot &\cdot
&\cdot & &\cdot \cr \cdot &\cdot &\cdot & &\cdot  \cr
0&0&0&\cdots&1\cr1&0&0&\cdots&0
\end{pmatrix}.
\end{gather*}
 Let $D$ be a positive divisor on $\Sigma$:
 \begin{gather*}
 D = \sum_{i=1}^{n}\nu_i, \qquad \nu_{i}\in \Sigma.
 \end{gather*}
 (Repetitions are allowed among the
 $\nu_i$.) We set $\tilde D$ to be the lift of this divisor to $\bbc$,
 under the projection $\rho$; this is an inf\/inite, but locally f\/inite,
 sum of points.
 Our phase space $M^{D}$ will be the product of the curve with the space of
 meromorphic functions on $\bbc$
 with values in $Gl(r,\bbc)$, having poles bounded by $\tilde D$
and  satisfying some quasiperiodicity relations:
 \begin{gather}
M^{D} = \Sigma \times \big\{ \phi\mid \phi \text{ is a meromorphic
} Gl(r,\bbc)\text{ --- valued function on }\bbc \text{ such
that}\nonumber
\\ \phantom{M^{D} =}{}
\text{ polar divisor }(\phi)\le \tilde D \text{ and }
\phi(\lambda+\omega_i) = I_i\phi (\lambda)I_i^{-1},\ i=1,2\big\}.
\label{mdee}
\end{gather}
This phase space comes equipped with a quadratic bracket def\/ined
in terms of an elliptic \mbox{$r$-matrix} \cite{sklyanin,
sklyanin2}. The Hamiltonians, as mentioned above, are the
coef\/f\/icients for the equation of the spectral curve of $\phi$.

The phase space $M^{D}$  is of the type we have been considering,
though this is not evident at f\/irst sight. The key is a way of
def\/ining vector bundles over curves which goes back to Weil. In
essence, if one has a f\/lat degree zero bundle over the curve,
sections of this bundle are given by sections of the trivial
bundle over the universal cover satisfying appropriate automorphy
relations, given by the holonomy matrices of the connection, which
then form a representation of the fundamental group. In the case
of an elliptic curve, the representation would be given by two
commuting matrices. This is not the case here, as the commutator
of $I_1, I_2$ is a root of unity. However, a generalisation of
Weil's procedure, given for example in \cite{Gro}, allows one to
construct bundles of non-zero degree; one takes representations
not of the fundamental group but of a central extension of the
fundamental group by appropriate roots of unity. For a degree one
bundle, the relation required is the one satisf\/ied by $I_1$,
$I_2$. The bundle obtained in this way is stable. By a theorem of
Atiyah \cite{atiyah}, all stable bundles on the elliptic curve
$\Sigma$ of degree one, rank $r$  are obtained from this one by
tensoring by a line bundle $V$ of degree zero, so that the moduli
space of stable bundles of degree one, rank $r$ is in fact
parametrised by the line bundles (mod, in fact, the $r$-th roots
of unity) and the parameter space is simply $J(\Sigma)= \Sigma$.

We have def\/ined then a bundle $E$ whose sections are given over
$\bbc$ as vector valued functions~$s$ satisfying the automorphy
relations:
\begin{gather*}
s(z+\omega_{i}) = I_{i}s(z).
\end{gather*}
This is not possible if one asks for the functions to be
holomorphic on all of~$\bbc$. One must allow some singularities;
one  chooses a point $p$ in $\Sigma$, and considers all of its
inverse images $\tilde p$ in~$\bbc$;  holomorphic sections of $E$
correspond to
 functions $s$ of the
form $z^{{-1}/{r}}\cdot$(holomorphic) near the points $\tilde p$,
and otherwise holomorphic on the plane; here $z$ is some
coordinate with $z=0$ corresponding to a $\tilde p$). In the same
vein, sections of $\mathop{\rm End}(E)$ over $\Sigma$ correspond
to holomorphic matrix valued functions over $\bbc$ satisfying the
quasi periodicity conditions of (\ref{mdee}). (For the
endomorphisms, holomorphic sections now correspond to holomorphic
functions.) The second factor of (\ref{mdee}) is thus the space
$H^{0}(\Sigma, \mathop{\rm End}(E)(D))=H^{0}(\Sigma, \mathop{\rm
End}(E)\otimes K_{\Sigma}(D))$ of meromorphic sections with poles
at $D$. (Recall that the bundle of one-forms is trivial on an
elliptic curve.)  As tensoring $E$ by a line bundle leaves the
automorphism bundle unchanged, one has that the space $  M^{D}$ is
a moduli space of pairs $(E\otimes V,\phi)$:
\begin{gather*}
M^{D} = \{(E',\phi)\mid E' \text{ stable},\ \mathop{\rm rank}r,
\text{ degree one over } \Sigma,\ \phi \in H^0(\Sigma, \mathop{\rm
End}(E)(D))\}.
\end{gather*}
{\it A basic section}. One wants to extract from a pair $(E'
=E\otimes V, \phi)$ a divisor $D_{L}$. We recall that $L$ was
def\/ined as the cokernel of the map
\begin{gather}
\phi-\xi\bbi\colon  E\otimes V (-D) \rightarrow E\otimes V.
\label{coker}
\end{gather}
One wants the divisor def\/ined by the vanishing of a suitable
section of $L$.

Sections $s = s(z) $ of $E\otimes V$ map to sections of $L$, by
the map (\ref{coker}), and as sections of $L$ they vanish when as
sections of $E\otimes V$ they lie in $(\phi-\xi\bbi)(E\otimes V
(-D) )$. Generically, lying in the image of $(\phi-\xi\bbi)$ is
the same as lying in the kernel of the matrix of cofactors:
\begin{gather*}
(\phi-\xi\bbi)_{\rm adj}\cdot s = 0.
\end{gather*}
This equation will be satisf\/ied at $g$ points $(z_{\mu},
\xi_{\mu})$, and it is these points we want in terms of $\phi$.
For this, we simply need to compute $s$.

In fact, it suf\/f\/ices to compute when $V$ is the trivial
bundle. The reason is that translation $T_{z_{0}}$ by $z_{0}$
along the curve has the same ef\/fect on $E$ as tensoring by a
line bundle, so that if $(z_{\mu}, \xi_{\mu})$ are coordinates for
$(E, \phi)$, then $(z_{\mu}- z_{0}, \xi_{\mu})$ are coordinates
for $(T_{z_{0}}^{*}(E), \phi)$. A consequence of this, by the way,
is that taking the reduction of (\ref{mdee}) to $Sl(r)$ is
equivalent to dropping the f\/irst factor and asking that the
matrices in the second factor belong to $Sl(r,\bbc)$. The
coordinates $(z_{\mu}, \xi_{\mu})$ then get shifted to their
`centre of mass':
\begin{gather*}
(z_{\mu}, \xi_{\mu})\mapsto \Biggl(z_{\mu} -
g^{-1}\sum_{\nu}z_{\nu},
\frac{\xi_{\mu}}{\prod_{\nu}\xi_{\nu}}\Biggr).
\end{gather*}
Returning to $Gl(r)$, to compute we need a section $s(z) =
(s_{1}(z), s_{2}(z),\dots,s_{r}(z))$ of $E$, expressed explicitly.
This can be done as follows, following \cite {HK-generalised}. We
normalise the periods of the elliptic curve $\Sigma$ to
$\omega_{1}= {1}/{r}, \omega_{2}= {\tau}/{r}$. Let us take the
$r$-th powers $f_{i}= s_{i-1}^{r}$ of the components of $s$, so
that
\begin{gather*}
s_{i}= f_{i+1}^{{1}/{r}}.
\end{gather*}
Let us `puncture' the curve $\Sigma$ at the point $((1+\tau)/2r)
$, and puncture the plane at the inverse images $((1+\tau)/2r) +
(1/r)\bbz+(\tau/r)\bbz$ of this point. We would then like  to
f\/ind an $r$-tuple $F$ of functions $(f_0,\dots , f_{r-1})$,
which are of the form $\zeta^{-1}({\rm holomorphic})^r$ near the
punctures (where $\zeta$ is a coordinate centered on the
puncture), satisfy
\begin{gather}
F\biggl(z+\frac1r\biggr) = F(z),\qquad
F\biggl(z+\frac{\tau}r\biggr) = I_2\cdot F(z), \label{period}
\end{gather}
and are such that the $r$-th roots along the real and imaginary
axes satisfy
\begin{gather}
(f_i)^{1/r}\biggl(z+\frac{1}{r}\biggr) = q^i (f_i)^{1/r} (z)\qquad
(f_i)^{1/r} \biggl(z+\frac{\tau}{r}\biggr) = (f_{i+1})^{1/r}(z).
\label{roots}
\end{gather}
Since $I_1^r= I_2^r = 1$, one is dealing with functions over the
elliptic curve $\Sigma' = \bbc/(\bbz\oplus\tau\bbz)$; let $\theta$
be the standard theta function for this curve; recall that it has
a simple zero at the points $((1+\tau)/2) + \bbz+\tau\bbz $, and
is otherwise non-zero and holomorphic. We distinguish two cases:

\noindent{\it Case 1:  $r$ is odd}.

Let
\begin{gather*}
\theta_{k,j}(z) = \theta\biggl(z + \frac{k+j\tau}{r}\biggr),\qquad
0\le k,j\le (r-1).
\end{gather*}
We have the relations:
\begin{gather*}
\theta_{k,j}(z+m) =\theta_{k,j}(z),\qquad \theta_{k,j}(z+m\tau)
=\exp\biggl(-\pi i m^2\tau-2\pi im\biggl(z+ \frac{k+j\tau
}{r}\biggr)\biggr)\theta_{k,j}(z),
\\
\theta_{k,j}\left(z + {\frac{1}{r}}\right) =\theta_{k+1,j}(z), \qquad
\theta_{k,j}(z + \frac{\tau}{r}) =\theta_{k,j+1}(z),\qquad 0\le
j<(r-1),
\\
\theta_{k,r-1}\left(z + \frac{\tau}{r}\right) =\theta_{k,0}(z)
\exp\biggl(-\pi i \tau-2\pi i \biggl(z +
\frac{k}{r}\biggr)\biggr),
 \label{theta}
\end{gather*}
where $m$ is an integer. Now if
\begin{gather*}
\rho_j = \frac{r-1}{2}-j,
\end{gather*}
we set
\begin{gather*}
f_j(z)  = \exp \biggl(2\pi i \tau\biggl(\frac{-jr(r-1)}{2} +
    \frac{(r-1)j(j+1)}{2}\biggr)\biggr)
\prod_{k=0}^{r-1}\left(
\frac{\theta_{k,j}^{r-2}(z)\theta_{k,j}(z+\rho_j\tau)}
{\prod\limits_{\ell=0,\ell\neq j}^{r-1}\theta_{k,\ell}(z)}\right).
\end{gather*}
Using the relations given for the $\theta_{k,j}$, one checks that
it has the
 correct form near the punctures, and that (\ref{period}) holds.
Now let $\tau$ be imaginary.  Let us consider the involutions
$f(z)\mapsto f(-z)$, $f(z) \mapsto \overline{f(\overline {z})}$.
Both these involutions preserve the poles and zeros of $f_0$. From
this, one has that $f_0$ must be even, as $f_0(0)\neq 0$. Using
the second involution, one can then multiply $f_0$ by a constant
$c$ so that $cf_0(0)$ is real. The function is then real on both
imaginary and real axes, and has no zeros on the axes.
From this, one has that  (\ref {roots}) holds for $f_0$. From the relations (\ref{theta}),
(\ref {roots}) follows for the other $f_i$. Deforming $\tau$, the
same then must hold for arbitrary $\tau$.

{\it  Case 2: $r$ is even.}

We then set
\begin{gather*}
\xi_{k,j}(z) = \theta\biggl(z + \frac{2k-1+2j\tau -\tau}{2r}
\biggr),\qquad 0\le k,j \le (r-1).
\end{gather*}
We have the relations:
\begin{gather*}
\xi_{k,j}(z\!+\!m)\!=\!\xi_{k,j}(z),\quad \xi_{k,j}(z+m\tau)\! =\!
\exp\biggl(\!-\pi i m^2\tau\!-\!2\pi im
\biggl(\!z\!+\!\frac{2k-1+2j\tau \!-\!\tau}{ 2r}
\biggr)\biggr)\xi_{k,j}(z),
\\
\xi_{k,j}\left(z + \frac{1}{r}\right)= \xi_{k+1,j}(z),\qquad
\xi_{k,j}\left(z + \frac{\tau}{ r}\right)= \xi_{k,j+1}(z),\qquad
0\le j<(r-1),
\\
\xi_{k,r-1}\left(z + \frac{\tau}{r}\right)=
\xi_{k,0}(z)\exp\biggl(-\pi i \tau-2\pi i \biggl(z+ \frac{2k
-1-\tau }{2r}\biggr)\biggr),
\end{gather*}
where $m$ is an integer. We then def\/ine
\begin{gather*}
\rho_{j} = \frac{r}{ 2} - j
\end{gather*}
and set{\samepage
\begin{gather*}
f_j(z)  = (-1)^j\exp \biggl(2\pi i \tau\biggl(\frac{-j(r-1)}{ 2}
\biggr)\biggr)
\prod_{k=0}^{r-1}\left(\frac{\xi_{k,j}^{r-1}(z)\xi_{k,j}(z+\rho_{j}\tau)
}{\prod\limits_{\ell=0}^{r-1}\xi_{k,\ell}(z)}\right).
\end{gather*}
Again, the $r$-th roots of the $f_j$ def\/ine our section.

}

{\it Separating coordinates}. As explained above, when $V$ is
trivial, the separating coordinates on~$M^{D}$ corresponding to an
element $\phi$ are given by the solutions $(z_\mu,\xi_\mu)$ of
\begin{gather*}
(\phi(z)-\xi\bbi)_{\rm adj}\cdot s = 0.
\end{gather*}
For $V$ non trivial and represented by the translation $z_0$ on
the curve, the coordinates are \mbox{$(z_\mu-z_0,\xi_\mu)$}.

{\it The symplectic form}. As outlined above, the symplectic form,
expressed in the divisor coordinates, is inherited from the
surface $X$; if $(z, \xi)$ are the coordinates def\/ined as above
on the surface, and our symplectic form on the surface is
$f(z,\xi) dz\wedge d\xi$, then the symplectic form in terms  of
the divisor coordinates is $\sum_{\mu}f(z_{\mu},\xi_{\mu})
dz_{\mu}\wedge d\xi_{\mu}$. In our case, $f(z,\xi)  = {1}/{\xi}$.

{\it The separation of variables}. The actual separation of
variables in terms of our Hamiltonians is given by a simple
generating function. Choose a basis $H_{1},\dots,H_{g}$ for the
space of Hamiltonians; one has in addition a certain number of
Casimirs, which we label as $H_{g+1},\dots,H_{k}$. Fixing
the~$H_i$~f\/ixes the spectral curve, and so determines $\xi$ as a
function of $z$:
\begin{gather*}
\xi = \xi(z, H_{1},\dots,H_{k}).
\end{gather*}
Choosing a base point on the spectral curve, we set
\begin{gather*}
F(z_{1},\ldots,z_{g}, H_{1},\dots,H_{k}) =
\sum_{\mu}\int_{z_{0}}^{z_{\mu}}\ln (\xi(z,
H_{1},\dots,H_{k}))\,dz.
\end{gather*}
The linearising coordinates of the f\/lows are given by
\begin{gather*}
Q_{i} = \frac{\partial F}{\partial H_{i}} =
\sum_{\mu}\int_{z_{0}}^{z_{\mu}}\xi^{-1}\frac{\partial
\xi}{\partial H_{i}} dz,\qquad i=1,\dots, g.
\end{gather*}
One can show that these are sums of Abelian integrals.
\section{Starting from surfaces}

One can invert the procedure outlined in the preceding section,
and start from a symplectic surface, or indeed a Poisson surface,
and create an integrable system; these systems, in some sense,
arrive already separated. This approach is in fact quite fruitful,
and has been explained by Vanhaecke in two interesting papers
\cite{Vh1,Vh2}.

The ingredients in general are a Poisson surface $X$ (af\/f\/ine
or projective) and a family of curves~$S_b$, $b\in B$ on the
surface. We ask that  $B$ map surjectively to a variety $C$, whose
functions will be  the Casimirs of the construction. One then has
an induced Poisson structure on the f\/iberwise $d$-th symmetric
product over $B$ of the family of curves, where $d$ is the
dif\/ference of dimensions between $B$ and $C$. The symplectic
leaves are then the f\/ibers over $C$.  Already, the construction
yields many interesting examples with $X = \bbc^2$.

One can restrict  a bit more the geometry of the construction. Let
$V$ be the divisor over which the Poisson tensor of the surface
vanishes. Let us suppose that the surface is projective, that the
system of curves has no base points, and let $C$ be the variety
describing the intersection of the curves with $V$, so that the
family of curves parametrised in $B$ as the inverse image of any
given point in $C$ have f\/ixed intersection in $X$ with $V$. One
can then show that the f\/ibers of $B$ over $C$ are
$g$-dimensional, where $g$ is the genus of the generic curve of
the family, and that the normal vector f\/ields describing
variations of curves within these f\/ibers are the images under
the Poisson tensor of holomorphic forms on the curves. When one
goes to the ($g$-dimensional) f\/iberwise symmetric product of the
curves over $B$, one obtains a Poisson manifold which is
birational to the family of Jacobians.

The family of projective Poisson surfaces is unfortunately a bit
limited, mainly because such surfaces must have Kodaira dimension
$0$ or $-\infty$. Noting that Poisson structures map well under
blowing down, one is reduced essentially to Abelian surfaces, K3
surfaces, (both symplectic), ruled surfaces and the projective
plane. One can show that the ruled surface over a curve $\Sigma$,
with a few exceptions, must be of the form $\bbp({\mathcal
O}\oplus {\mathcal K}(D))$, where $D$ is a positive divisor on the
curve. These, in essence, are already covered by the cases
outlined above, in Section 3.

\section{Comments}

\subsection{Multi-Hamiltonian structures}

The picture presented above, of decomposition into symmetric
products, is also natural in the context of multi-Hamiltonian
structures. The key point here is that any linear combination of
Poisson structures in two dimensions is a Poisson structure. This
principle propagates to symmetric products and gives a
multi-Hamiltonian structure on the system. This gives one way of
approaching  multi-Hamiltonian structures examined in a certain
number of recent papers on particular cases of the systems
discussed above (e.g.~\cite{BFP, FP}).

For example, if one considers the space $M_{{\rm rat},n}$ of
matricial polynomials of degree at most $n$, one can build an
$n+3$ dimensional space of Poisson structures on this space which
incorporates both the rational linear $r$-matrix bracket and the
rational quadratic (Sklyanin) bracket. These Poisson structures
are induced in the way outlined in Section~3 from an  $n+3$
dimensional family of Poisson structures on the total space of the
line bundle ${\mathcal O}(n)$. In the tensor bracket notation of
\cite {FT}:
\begin{gather*}
\{\phi(\lambda) {\stackrel {\otimes} ,}\phi(\mu)\}_{a,b} :=
\biggl[r(\lambda-\mu), \phi(\lambda) \otimes \biggl(a(\mu)\bbi
-\frac{b}{2} \phi(\mu)\biggr) + \biggl(a(\lambda)\bbi -\frac{b}{2}
\phi(\lambda)\biggr)\otimes \phi(\mu)\biggr].
\end{gather*}

Here $a(\lambda) $ is an arbitrary polynomial of degree at most
$(n+1)$,  and $b$ is an arbitrary constant.

The spectral curve procedure outlined above gives us a spectral
curve in the total space $X$ of the line bundle ${\mathcal O}(n)$
over $\bbp^{1}$. The space $X$ has a family of Poisson structures
given in standard coordinates by $(a(z)
+b\xi)({\partial}/{\partial z}\wedge {\partial}/{\partial \xi})$,
and a result of \cite{HaH-multihamiltonian} tells us that these
structures correspond to the family of brackets given above under
the passage to divisor coordinates. In fact, the coordinates
$(z_{\mu},\xi_{\mu})$ are Nijenhuis coordinates in a suitable
sense for this family of Poisson structures
\cite{HaH-multihamiltonian}.

The same holds also in the elliptic and trigonometric cases;
again, see \cite{HaH-multihamiltonian}.

We should note that the existence of multi-Hamiltonian structures
does not always imply that individual vector f\/ields are
multi-Hamiltonian, i.e., that for the f\/ield $V$ there are
functions~$f_i$ such that $V$ is generated by $f_i$ under the
$i$th Poisson structure. This question is explained
in~\cite{Vh1,Vh2}.

\subsection{Other groups and Prym varieties} As  we noted above,
integrable systems of the types listed above exist for complex
reductive groups other than $Gl(r)$. They do not quite f\/it into
our set-up, as they are not systems of Jacobians, but rather of
Prym varieties. Indeed, the construction we gave relied heavily on
the existence of a spectral curve, that is a curve of eigenvalues
of $\phi$. The analogous object in the general case, as developed
in \cite {donagi, faltings, scognamillo} is the curve $C$ of
conjugates of $\phi$ in the Lie algebra $\gh$ of a maximal torus
$H$. This curve is Galois over the base curve, with Galois group
the Weyl group of $G$. The analog of a line bundle in turn is a
Weyl-invariant bundle $P_{H}$ with $H$ as structure group. Let
${\mathcal L}$ denote the integer lattice in $\gh$. The analogue
of the Jacobian, classifying invariant $H$-bundles, is given by
\begin{gather*}
( J(C)\otimes_{Z}{\mathcal L})^{W}.
\end{gather*}
The connected component of the identity is known as a generalised
Prym variety. The Lag\-rangian leaves of our systems are of this
type.

There is for integrable systems of Prym varieties a corresponding
notion of being of rank~2; one obtains, not a surface $Q$, but a
variety $P$ of dimension $r+1$, where $r = \dim (H)$ \cite
{HM-pryms}. This variety, and the family of curves, also encodes
the system. The cases  of the examples cited above, generalised to
arbitrary $G$ are also of rank 2 in this generalised sense. There
are no longer natural separating coordinates, but one can perform
the actual integration to obtain variables~$Q_{i}$.

Indeed, choosing a suitable basis $e_{i}$ for $\gh $, $H$-bundles
decompose as a sum of line bundles~$L_{i}$; one can for each of
these obtain divisor functions $z_{i,\mu}$, $\xi^{i}_{\mu}$, where
$\xi^{i}$ is the coordinate of the spectral curve (which lies in
$K(D)\otimes \gh$) in the direction $e^{i}$ dual to $e_{i}$. These
functions no longer form a coordinate system, as there is some
redundancy due to the $W$-invariance. The symplectic form is still
$\sum_{{i,\mu}}f(z_{i,\mu}, \xi^{i}_{\mu}) dz_{i,\mu}\wedge
d\xi^{i}_{\mu}$, though, and writing $\xi$ as a function of $z$
and the independent Hamiltonians $H_{i}$ as above, def\/ining the
same generating function, as above, gives again the dual variables
$Q_{i}$.

\pdfbookmark[1]{References}{ref}
\LastPageEnding

\end{document}